\documentstyle[11pt,newpasp,twoside,psfig]{article}
\newcommand{\eqn}[1]{equation~(\ref{eqn:#1})}

\newcommand{\mbf}[1]{\mbox{\boldmath $#1$} }

\newcommand{\trace}{{\mathrm{Tr}}}
\newcommand{\pauli}[1]{{\mbf{\sigma}}_{#1}}

\newcommand{\psr}{PSR\,J0437$-$4715}

\begin{document}
\title{Polarimetric Pulse Profile Modeling:
Applications to High-Precision Timing and Instrumental Calibration }
\author{Willem van Straten}
\affil{Netherlands Foundation for Research in Astronomy, \\
P.O. Box 2, 7990 AA Dwingeloo, The Netherlands}

\begin{abstract}
A new method is presented for modeling the transformation between two
polarimetric pulse profiles in the Fourier domain.  In practice, one
is a well-determined standard with high signal-to-noise ratio and the
other is an observation that is to be fitted to the standard.  From
this fit, both the longitudinal shift and the polarimetric
transformation between the two profiles are determined.  Arrival time
estimates derived from the best-fit longitudinal shift are shown to
exhibit greater precision than those derived from the total intensity
profile alone.  In addition, the polarimetric transformation obtained
through this method may be used to completely calibrate the
instrumental response in observations of other sources.
\end{abstract}

\section{Polarimetric Pulse Profile Modeling}
\label{sec:modeling}

In radio astronomy, a pulsar's mean polarimetric pulse profile is
measured by averaging the observed Stokes parameters as a function of
pulse longitude.  By integrating many well-calibrated pulse profiles,
a standard profile with high signal-to-noise ratio (SNR) may be formed
and used as a template to which individual observations are fit.  For
example, in Appendix A of Taylor (1992), a method is presented for
modeling the relationship between standard and observed total
intensity profiles in the Fourier domain.  In the current treatment,
the scalar equation that relates two total intensity profiles is
replaced by an analogous matrix equation, which is expressed using the
Jones calculus.  The polarization of the electromagnetic field is
described by the coherency matrix,
${\mbf\rho}=(S_0\,\pauli{0}+\mbf{S\cdot\sigma})/2$, where $S_0$ is the
total intensity, $\mbf{S} = (S_1,S_2,S_3)$ is the Stokes polarization
vector, $\pauli{0}$ is the $2\times2$ identity matrix, and
$\mbf{\sigma} = (\pauli{1},\pauli{2},\pauli{3})$ are the Pauli spin
matrices (Britton 2000).  Under a linear transformation of the
electric field vector as represented by the Jones matrix, ${\bf J}$,
the coherency matrix is subjected to the congruence transformation,
${\mbf{\rho}^\prime}={\bf{J}}\mbf{\rho}{\bf{J}}^\dagger$ (Hamaker
2000).

Let the coherency matrices, $\mbf{\rho}^\prime (\phi_n)$, represent
the observed polarization as a function of discrete pulse longitude,
$\phi_n$, where $0\le n< N$ and $N$ is the number of pulse longitude
intervals.  Each observed polarimetric profile is related to the
standard, $\mbf{\rho}_0(\phi_n)$, by the matrix expression,
\begin{equation}
\label{eqn:model}
\mbf{\rho}^\prime (\phi_n) = \mbf{\rho}_{\mathrm DC}
	+ \mbf{\rho}_{\mathrm N}(\phi_n)
	+ {\bf J} \mbf{\rho}_0 (\phi_n - \varphi) {\bf J}^\dagger,
\end{equation}
where $\mbf{\rho}_{\mathrm DC}$ is the DC offset between the two
profiles, $\mbf{\rho}_{\mathrm N}$ represents the system noise, ${\bf
J}$ is the polarimetric transformation and $\varphi$ is the longitudinal
shift between the two profiles.  The Jones matrix, ${\bf J}$, is
analogous to the gain factor, $b$, in equation (1) of Taylor (1992).
However, as ${\bf J}$ has seven non-degenerate degrees of freedom, the
matrix formulation introduces six additional free parameters.  The
discrete Fourier transform (DFT) of \eqn{model} yields
\begin{equation}
\mbf{\rho}^\prime (\nu_m) = \mbf{\rho}_{\mathrm N}(\nu_m) + {\bf J}
  \mbf{\rho}_0 (\nu_m) {\bf J}^\dagger \exp(-i2\pi m \varphi),
\label{eqn:fourier_rho}
\end{equation}
where $\nu_m$ is the discrete pulse frequency.  Given the measured
Stokes parameters, $S_k^\prime(\phi_n)$, and their DFTs,
$S_k^\prime(\nu_m)$, the best-fit model parameters will minimize 
the objective merit function,
\begin{equation}
\chi^2 = \sum_{m=1}^{N/2} \sum_{k=0}^3 { |S_k^\prime(\nu_m) -
\trace[\pauli{k}\;\mbf{\rho}^\prime(\nu_m)]|^2 \over \varsigma_k^2 },
\label{eqn:merit}
\end{equation}
where $\varsigma_k$ is calculated from the noise power and $\trace$ is
the matrix trace.  As in van Straten (2004), the partial derivatives
of \eqn{merit} are computed with respect to both $\varphi$ and the
seven parameters that determine ${\bf J}$.  The Levenberg-Marquardt
method is then applied to find the parameters that minimize $\chi^2$.


\section{Applications}

Dual-polarization observations of \psr\ were made using the Parkes
Multibeam receiver and CPSR-II, the 128\,MHz baseband recording and
real-time processing system at the Parkes Observatory.  The data were
observed during seven separate sessions between 5~June and
21~September~2003, calibrated using the method described in van
Straten (2004), and integrated to produce the polarimetric standard
shown in Figure~1.


\begin{figure}[h]
\label{fig:std}
\centerline{\psfig{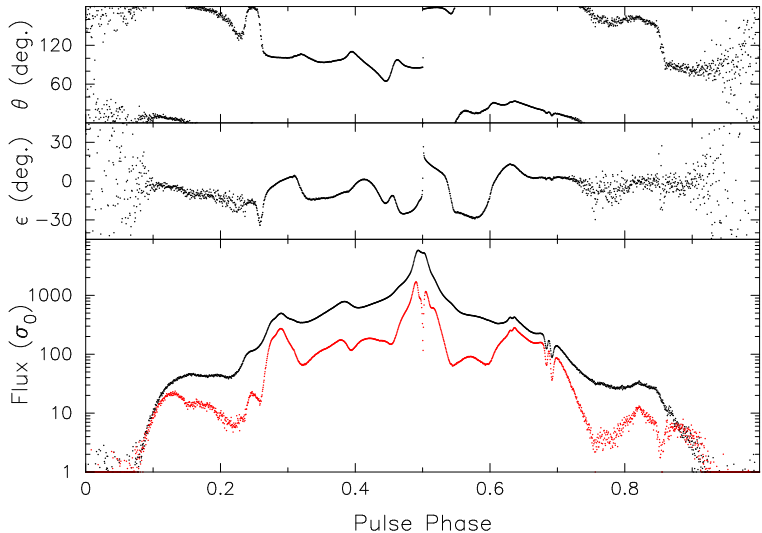}}
\caption{Mean polarization of \psr, plotted using polar coordinates:
orientation, $\theta$; ellipticity, $\epsilon$; and polarized
intensity, $S=|\mbf{S}|$ (plotted in red below the total intensity).
Flux densities are normalized by $\sigma_0$, the r.m.s.\ of the
off-pulse total intensity phase bins.  Data were integrated over a
64\,MHz band centered at 1341\,MHz for approximately 40.6~hours.}
\end{figure}

\subsection{High-Precision Timing}

In any high-precision pulsar timing experiment, the confidence limits
placed on the derived physical parameters of interest are proportional
to the precision with which pulse time-of-arrival (TOA) estimates can
be made.  Aside from typical observational constraints such as system
temperature, instrumental bandwidth, and allocated time, TOA precision
also fundamentally depends upon the physical properties of the pulsar,
including its flux density, pulse period, and the shape of its mean
pulse profile.  When fully resolved, sharp features in the mean pulse
profile generate additional power in the high frequency components of
its Fourier transform.  As higher frequencies contribute stronger
constraints on the linear phase gradient in the last term of
\eqn{fourier_rho}, sharp profile features translate into greater
arrival time precision.

This important property may be exploited in order to significantly
improve the precision of arrival time estimates derived from full
polarimetric data.  As noted by Kramer et al.\ (1999), the mean
profiles of Stokes~$Q$, $U$, and~$V$ may contain much sharper features
than that of Stokes~$I$, especially when the pulsar exhibits
transitions between orthogonally polarized modes.  Although the
polarized flux density is lower than the total intensity, these
transitions lead to a greater SNR in the high frequency components of
the polarized power spectra, as shown in Figure~2.

\begin{figure}
\label{fig:snr}
\centerline{\psfig{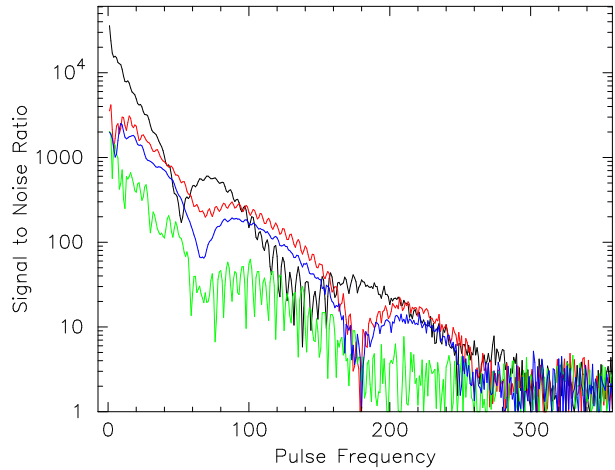}}
\caption{Power spectra of the standard profile plotted in Figure~1.
Stokes~$I$, $Q$, $U$, and~$V$ are plotted in black, red, green, and
blue.  Above 40 times the pulse frequency, both Stokes~$Q$
and~$V$ exhibit a SNR greater than or equal to that of the total
intensity.}
\end{figure}

To demonstrate the potential for improved arrival time precision, TOA
estimates spanning over 100 days were derived from the 490 five-minute
integrations used to produce the standard plotted in Figure~1.  The
arrival times derived from the mean total intensity profile have a
post-fit residual r.m.s.\ of 198\,ns.  By modeling the polarimetric
pulse profile as described in Section~1, the resulting arrival times
have a post-fit residual r.m.s.\ of only 146\,ns, an improvement in
precision of approximately 36\%.

\subsection{Instrumental Calibration}

Based on the assumption that the mean polarimetric pulse profile does
not vary significantly with time, the standard profile and modeling
method may also be used to determine the polarimetric response of the
observatory instrumentation at other epochs.  As a demonstration, a
single, uncalibrated, five-minute integration of \psr\ was fitted to
the polarimetric standard shown in Figure~1.  The model was solved
independently in each of the 128 frequency channels, producing the
instrumental parameters shown with their formal standard deviations in
Figure~3.  In each 500\,kHz channel, it is possible to estimate the
ellipticities and orientations of the feed receptors with an
uncertainty of only one degree.  This unique model of the instrumental
response may be used to calibrate observations of other point sources.

\begin{figure}[h]
\label{fig:fit}
\centerline{\psfig{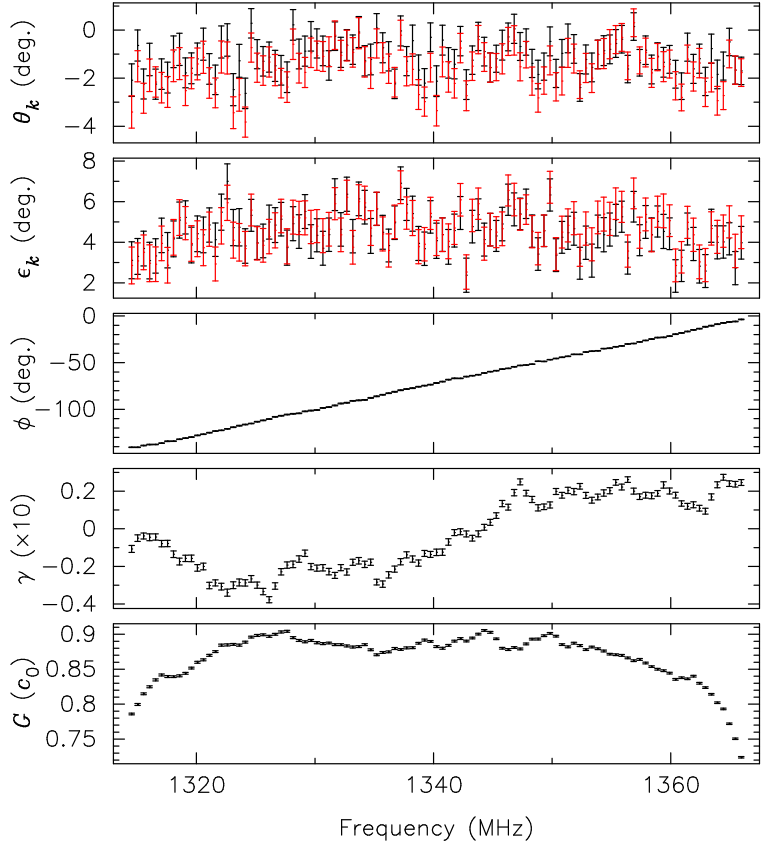}}
\caption{Best-fit instrumental parameters as a function of observing
frequency.  A single, uncalibrated, five-minute integration of \psr\
was fitted to the polarimetric standard. From top to bottom are
plotted the orientations, $\theta_k$, and ellipticities, $\epsilon_k$,
of the feed receptors, the differential phase, $\phi$, the
differential gain, $\gamma$, and the absolute gain, $G$, specified in
units of an intermediate reference voltage that can later be
calibrated to produce absolute flux estimates.  In the top two panels,
black and red correspond to receptors 0 and 1, respectively.}
\end{figure}

\section{Conclusion}

When compared with the scalar equation used to model the relationship
between total intensity profiles, the matrix equation presented in
Section~1 quadruples the number of observational constraints while
introducing only six additional free parameters.  By completely
utilizing all of the information available in mean polarimetric pulse
profiles, arrival time estimates may be obtained with greater
precision than those derived from the total intensity profile alone.
In addition, the modeling method may be used to uniquely determine the
polarimetric response of the observatory instrumentation using only a
short observation of a well-known source.

\acknowledgments The Swinburne University of Technology Pulsar Group
provided the CPSR-II observations presented in this poster.  The
Parkes Observatory is part of the Australia Telescope which is funded
by the Commonwealth of Australia for operation as a National Facility
managed by CSIRO.


\begin{references}
\reference Britton,~M.~C., 2000, \apj, 532, 1240
\reference Hamaker,~J.~P., 2000, \aaps, 143, 515
\reference Kramer,~M., Doroshenko,~O., \& Xilouris,~K.~M., 1999,
poster presented at Pulsar Timing, General Relativity and the Internal
Structure of Neutron Stars
\reference van Straten,~W., 2004, \apjs, 152, {\it in press}
\reference Taylor,~J.~H., 1992, Phil. Trans. R. Soc. Lond. A, 341, 117
\end{references}
\end{document}